\documentclass[aps,prl,reprint,letterpaper,showpacs,amsmath,amssymb]{revtex4-1}
\usepackage{graphicx}
\usepackage[absolute]{textpos}

\begin{document}
\begin{textblock*}{8.5in}(0.1in,0.25in)
\begin{center}
PHYSICAL REVIEW B \textbf{92}, 045307 (2015)
\end{center}
\end{textblock*}
\begin{textblock*}{2.5in}(5.06in,2.77in)
{\small \doi{10.1103/PhysRevB.92.045307}}
\end{textblock*}

\title{Heating by exciton and biexciton recombination in GaAs/AlGaAs quantum wells}


\author{V.V. Belykh}
\email[]{belykh@lebedev.ru}
\thanks{present address: Experimentelle Physik 2, Technische Universit\"{a}t Dortmund, 44227 Dortmund, Germany}
\affiliation{P.N. Lebedev Physical Institute of the Russian Academy of
Sciences, Moscow, 119991 Russia}
\author{M.V. Kochiev}
\affiliation{P.N. Lebedev Physical Institute of the Russian Academy of
Sciences, Moscow, 119991 Russia}
\received{15 March 2015} \revised{21 May 2015} \published{27 July
2015}

\begin{abstract}
A comprehensive experimental investigation of exciton and biexciton
recombination in GaAs/AlGaAs quantum wells is presented. Exciton and
biexciton recombination times are found to be 16 and 55~ps,
respectively. A method of determining the dynamics of the exciton
temperature is developed. It is shown that both exciton and
biexciton recombination processes increase the exciton temperature
by an amount as high as $\sim 10$~K. These processes impose a new
restriction on the possibility of exciton Bose-Einstein condensation
and make impossible its achievement in a system of direct excitons
in GaAs quantum wells even for resonantly excited exciton gas.
\end{abstract}


\maketitle

\section{Introduction}
Excitons in semiconductors are a simple and very convenient analog of
an atomic system. Like atoms, excitons can form molecules (biexcitons
\cite{Miller1982}) and ions (trions \cite{Kheng1993}). The Rydberg
energy and Bohr radius of excitons are $\sim 10$~meV and $\sim
10$~nm as compared to $\sim 10$~eV and $\sim 0.1$~nm for atoms. This
makes excitons more susceptible to external fields and allows one to
implement in the laboratory conditions that are achievable only at
extreme places of the Universe for atoms. Examples include the
stabilization of the electron-hole liquid by a magnetic field
\cite{Kavetskaya1996} and spatial separation of electrons and holes
by an electric field in quantum wells (QWs), which makes excitons
indirect, suppressing the formation of exciton complexes and
enhancing exciton radiative lifetime \cite{Alexandrou1990,
Sivalertporn2012, Schinner2013}.

The achievement of Bose-Einstein condensation (BEC) in a system of
ultracold atoms two decades ago \cite{Anderson1995, Davis1995}
together with rapid progress in nanotechnology inspired the work
which has led to the attainment of BEC in a system of spatially
indirect excitons in QWs \cite{Butov2002, High2012, Gorbunov2012}
and exciton polaritons in microcavities \cite{Kasprzak2006,
Balili2007}. Atomic BEC was achieved at temperatures of $\sim
10^{-7}$~K, which comprises the main difficulty in the experiment.
On the other hand, the BEC of excitons can be reached at
temperatures of $\sim 1$~K due to their much smaller effective mass
compared to the atomic mass. Excitons inevitably interact with
thermal reservoir, the lattice, through the emission and absorption
of phonons, and exciton system cooling is accomplished via the
lattice cooling. On the other hand, an atomic system can be left on
its own in a magneto-optical trap without interaction with any
thermal reservoir, and the temperature of such a system is decreased
by laser and evaporative cooling techniques \cite{Cornell2002}.
Similarly to atomic BEC, the achievement of exciton BEC suffers from
inelastic exciton-exciton collisions leading to biexciton formation \cite{Miller1982, Lovering1992, Phillips1992, Kim1994, Kim1998, Mootz2014}
and even to collapse into electron-hole liquid
\cite{JeffriesC.D.1983, Burbaev2007, Bagaev2010}. The exciton system
in direct gap semiconductors suffers also from radiative
recombination. This is why exciton BEC was searched for indirect
excitons with suppressed recombination and Coulomb attraction.

So far, it was believed that the main effect of exciton radiative
recombination and biexciton formation on BEC is the decrease of
exciton population which can be faster than thermalization of the
system with the lattice \cite{High2012}. In this case, it would be
possible to attain BEC for resonant excitation of cold gas of direct
excitons within an exciton population lifetime which is $\sim 0.5$~ns
in GaAs QWs. This lifetime and even exciton recombination time
\cite{Deveaud1991} are longer than exciton thermalization time
\cite{Knox1986, Knox1988}. However, so far, only coherence mediated by the exciting
laser was observed for exciton gas at low densities \cite{Haacke1997, Garro1999, Langbein2000a, Hayes2000, Savona2000, Kocherscheidt2003, Savona2002}, and no signatures of spontaneous
coherence were reported for direct excitons, to the best of our
knowledge. In this paper, we show that exciton and biexciton
recombination leads to significant heating of the exciton system.
This imposes a new restriction on the possibility of exciton BEC. In
particular, recombination heating makes it impossible to achieve BEC
in a system of direct excitons in GaAs QWs even for resonantly
excited cold exciton gas.

First, we determine the exciton and biexciton recombination times in
QWs. In previous studies, the exciton recombination time (radiative
decay time of low-momenta radiative excitons) was extracted from the
photoluminescence (PL) kinetics
of resonantly created excitons 
using sophisticated models taking into account different scattering
mechanisms \cite{Deveaud1991, Vinattieri1994}. The exciton
recombination time was also determined in four-wave mixing
experiments \cite{Langbein2000} and calculated theoretically
\cite{Andreani1991}. Here, by using exciton-resonant excitation with
different polarizations and powers and also different temperatures,
we clearly show contributions of different dephasing processes to the
decay rate of the fast component in the PL kinetics and determine
the exciton recombination time. The biexciton recombination time so
far was determined only in four-wave mixing experiments
\cite{Langbein2000} and calculated theoretically
\cite{Citrin1994,Ivanov1997}. In this study, we determine biexciton recombination time directly by observing the
decay of biexciton PL and rise of exciton PL upon the resonant
excitation of biexcitons.

Heating by exciton recombination (evaporative optical heating) was
considered previously theoretically for free carriers
\cite{Bimberg1985} and indirect excitons \cite{Ivanov2004} in QWs.
However, to study this effect experimentally, one has to measure
fairly small temperature changes. Here, we develop and ground a
sensitive method to determine the exciton temperature dynamics based
on the temperature dependence of the exciton
population decay rate. 
We show both theoretically and experimentally that, for the
considered system of direct excitons, the exciton temperature is
higher than the lattice temperature by $\gtrsim 4$~K due to exciton
recombination heating. Furthermore, we demonstrate that the formation of biexcitons followed by their
recombination also contributes significantly to the heating of the
system. With increasing excitation power, the temperature increases
by an amount proportional to the ratio of the biexciton and exciton
concentrations, and this amount becomes as high as $\sim 10$~K.

\section{Experimental details}
The sample under study is a GaAs/Al$_{0.05}$Ga$_{0.95}$As
heterostructure with two shallow tunneling-isolated QWs of widths
$d=3$ and 4 nm. The thickness of the Al$_{0.05}$Ga$_{0.95}$As
barrier layer separating the QWs is 60~nm. Only states of the wider
QW were excited and studied. The same sample was used in
Ref.~\cite{Kochiev2012}, where the formation of trions as a result
of the preferential capture of one species of carriers into the QWs
under above-barrier excitation was studied.

The sample is mounted in a He-vapor optical cryostat and excited by
the radiation of a mode-locked Ti-sapphire laser generating a
periodic train of 2.5-ps-long pulses at a repetition rate of 76~MHz.
The excitation laser beam is focused into a 10- to 20-$\mu$m spot on
the sample surface using a 6-mm-focus micro-objective located in
front of the sample surface so that the surface is near its focal
plane. The PL is collected by the same micro-objective. The
excitation beam was slightly misaligned with respect to the optical
axis, which allows to block the direct reflection. The
micro-objective is mounted on the sample holder inside the cryostat,
which provides the good stability of the system against vibrations.
The PL coming out from the cryostat is focused with a $\sim
100$-mm-focus lens to form a magnified image of the PL spot on the
slit of a spectrometer coupled to a Hamamatsu streak camera. The
slits of the spectrometer and streak camera selected the central
region of the PL spot with homogeneous intensity distribution. In
all experiments except those described in Section~\ref{SecExRec}
(Fig.~\ref{ExDec}), the excitation beam was linearly polarized and
PL was registered in a perpendicular linear polarization. The spectral resolution is $0.2-0.5$~meV. The
temporal resolution is $5-20$~ps, depending on the spectral resolution and
the used time range.

\section{Results and discussion}
\begin{figure}
\begin{center}
\includegraphics[width=\columnwidth]{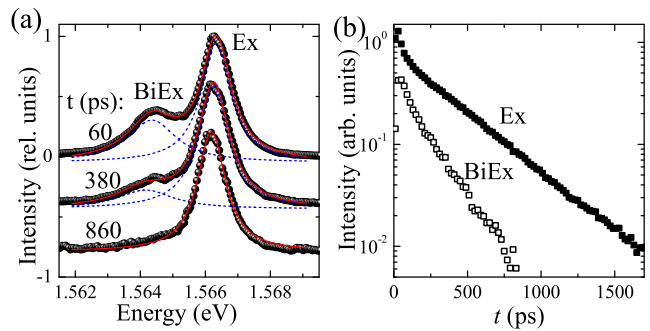}
\caption{(a) PL spectra, corresponding to the different times after
the excitation pulse (symbols) fitted with Lorentzian peaks (lines).
The spectra are normalized to the maximum value and vertically
shifted. (b) Dynamics of the exciton (full symbols) and biexciton
(open symbols) PL intensities. (a),(b) Exciton-resonant excitation
($\Delta=0$~meV) with $P = 0.25$~mW, $T_\text{latt}=10$~K. }
\label{Spectra}
\end{center}
\end{figure}

The QW emission spectra for different times $t$ after a resonant
excitation pulse are presented in Fig.~\ref{Spectra}(a). The spectra
feature two lines separated by about $2$~meV. The higher-energy line
(Ex) is attributed to exciton emission. It persists in the spectra
even for sufficiently high temperatures and low e-h densities. The
lower-energy line (BiEx) is significantly reduced with respect to
the exciton line as the e-h density decays with time
(Fig.~\ref{Spectra}(a)). We attribute this low-energy line to
biexciton emission. Indeed, generation of biexcitons is the most efficient for
the resonant excitation \cite{Lovering1992, Phillips1992}. The spectrum is fitted with two Lorentzian peaks
(lines in Fig.~\ref{Spectra}(a)) to determine the intensities,
widths, and spectral positions of the exciton and biexciton lines.
The intensity of the BiEx line decays about two times faster than
the intensity of the Ex line (Fig.~\ref{Spectra}(b)) as expected for
biexcitons \cite{Kim1994, Kim1998}, and the biexciton nature of the
BiEx line will be further confirmed throughout the paper.

\subsection{Exciton recombination time}
\label{SecExRec}
\begin{figure*}
\begin{center}
\includegraphics[width=1.8\columnwidth]{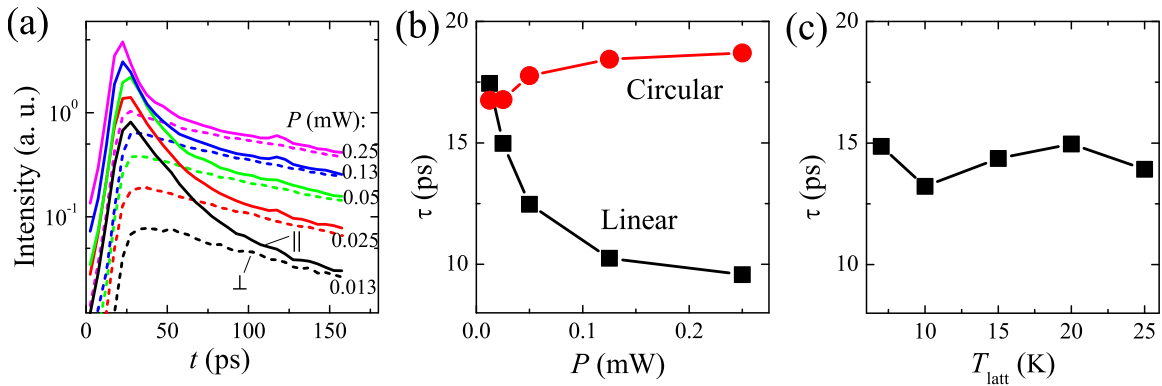}
\caption{Exciton-resonant excitation ($\Delta=0$ meV). (a) Exciton
intensity dynamics in linear polarizations parallel (solid lines)
and perpendicular (dashed lines) to the excitation polarization for
different excitation powers. $T_\text{latt}=10$~K. (b) Power
dependences of the fast component decay time for the linearly
(squares) and circularly (circles) polarized excitations.
$T_\text{latt} = 10$~K. (c) Temperature dependence of the fast
component decay time for linearly polarized excitation with
$P=0.013$~mW (different excitation spot position on the sample).} \label{ExDec}
\end{center}
\end{figure*}
Here we determine the exciton recombination time (radiative decay
time of low-momenta radiative excitons) by investigating the exciton
dynamics with temporal resolution of about 5~ps for
exciton-resonant linearly polarized excitation.
Figure~\ref{ExDec}(a) shows exciton intensity dynamics registered in
polarization parallel ($\parallel$, solid curves) and perpendicular
($\perp$, dashed curves) to the excitation polarization. Two decay
components are clearly seen in the $\parallel$ polarization
kinetics, while the $\perp$ polarization kinetics shows the slow
component only. We interpret these observations as follows. A
resonant excitation pulse creates a large population of low-$k$
radiative excitons with wave vectors within the light-cone.
Accordingly, the fast component is induced by the initial low-$k$
exciton recombination and scattering towards the nonradiative
reservoir, while the slow component corresponds to the decay of the
reservoir exciton population. The contribution of a spin relaxation
to the decay of the fast component is small \cite{Langbein2000}. Otherwise, the PL decay
in $\parallel$ polarization would be accompanied by the
comparable PL rise in the $\perp$ polarization, which is not seen
in the experiment (Fig.~\ref{ExDec}(a)). The initial scattering of
the low-$k$ excitons to the nonradiative states corresponds to the
increase of the mean exciton energy (temperature) either due to the
absorption of phonons (the corresponding scattering rate is
$1/\tau_\text{phon}$) or to the formation of biexcitons (the
corresponding rate is $1/\tau_\text{f}$). Indeed, each
biexciton formation event releases energy for the exciton system
approximately equal to the biexciton binding energy. For the decay
rate of the fast component neglecting the scattering back from the
reservoir we can write
$1/\tau=1/\tau_0+1/\tau_\text{phon}+1/\tau_\text{f}$. As the
excitation density $P$ is increased, the rate of biexciton formation
increases and $\tau$ should decrease, which we do observe in the
experiment (black squares in Fig.~\ref{ExDec}(b)). In the limit of
$P\rightarrow0$ we have $1/\tau = 1/\tau_0+1/\tau_\text{phon}$.
Interestingly, for circularly polarized excitation, the fast
component decay time for the co-polarized intensity is almost
independent of $P$. This can be explained by the fact that biexciton
formation is suppressed for circularly polarized exciton population
since total biexciton spin equals zero \cite{De-Leon2002,
Miller1982}.

An increase in the lattice temperature $T_\text{latt}$ should favor
phonon absorption and lead to an increase in $1/\tau_\text{phon}$,
causing a decrease in $\tau$. However, Fig.~\ref{ExDec}(c) shows
that $\tau$ is almost independent of $T_\text{latt}$, indicating
that $\tau_\text{phon} \gg \tau$. As we show in the following, interaction with
phonons is characterized by times of $\sim 100$~ps. Note also that at low $P$ secondary emission
can preserve laser-mediated coherence for a time comparable to $\tau_0$ \cite{Haacke1997, Garro1999, Langbein2000a, Hayes2000, Savona2000, Kocherscheidt2003, Savona2002}, which, of course, does not prevent exciton radiative decay at the rate $1/\tau_{0}$. Finally, at
small $P$, $\tau \approx \tau_0$ and we get the exciton
recombination time $\tau_0 \approx 16 \pm 2$~ps. This result is
consistent with the results of kinetic measurements of
Ref.~\cite{Deveaud1991} (10~ps for a 4.5-nm QW) and
Ref.~\cite{Vinattieri1994} (20~ps for an 8-nm QW), with four-wave
mixing studies of Ref.~\cite{Langbein2000} (13~ps for a 25-nm QW)
and theoretical calculations Ref.~\cite{Andreani1991} (25~ps for a
10-nm QW).

\subsection{Biexciton recombination time}
\begin{figure}
\begin{center}
\includegraphics[width=0.8\columnwidth]{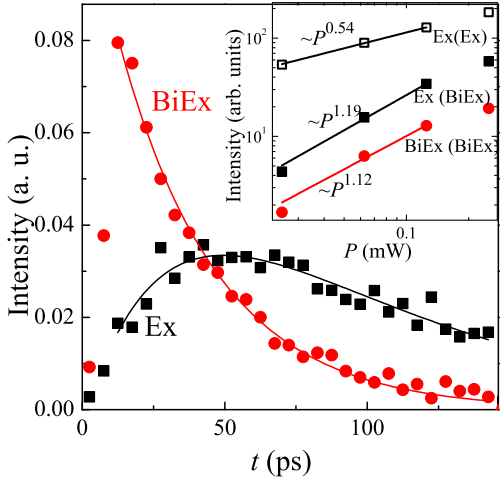}
\caption{Dynamics of exciton (squares) and biexciton (circles) PL
intensities for biexciton-resonant excitation ($\Delta=-1$ meV) with
$P = 0.13$~mW. The inset shows power dependences of the exciton
(squares) and biexciton (circles) time-integrated intensities for
biexciton-resonant (full symbols) and exciton-resonant (open
symbols) excitations. $T_\text{latt}=10$ K. } \label{BiExDec}
\end{center}
\end{figure}
\label{SecBiEx}

Now we determine the biexciton recombination time (radiative
lifetime) by investigating the exciton-biexciton dynamics with a
temporal resolution of about 5~ps for biexciton-resonant linearly
polarized excitation (Fig.~\ref{BiExDec}). Biexciton-resonant
excitation implies the absorption of two photons leading to the
creation of a biexciton having an energy of
$2E_\text{x}-E_\text{b}$. Thus, we used the excitation photon energy
$\hbar\omega=E_\text{x}-E_\text{b}/2$ detuned by $\Delta
=E_\text{laser}-E_\text{x}= -1$~meV from the exciton resonance.
Figure~\ref{BiExDec} shows, that for this excitation energy, the
biexciton intensity greatly exceeds the exciton intensity at short
times. Then, the biexciton intensity rapidly decays with a
characteristic time $\tau_1 \approx 35$~ps. The decay of the
biexciton intensity is accompanied by an increase in the exciton
intensity, which starts to decrease at later times.

The simplest scenario explaining the experimental observations is
the following. An excitation pulse creates biexciton population. In
a process of biexciton radiative decay, one of the e-h pairs
constituting a biexciton recombines to give a photon with energy
$E_\text{x}-E_\text{b}$ and leaves an exciton. Thus, the radiative
decay of biexciton population is accompanied by the creation of the
same amount of excitons. Subsequently excitons either decay
radiatively or form secondary biexcitons giving much weaker emission
at longer times, so the latter process can be neglected. According
to this scenario the time-integrated biexciton intensity should be
equal to the time-integrated exciton intensity. In the experiment,
however, the total exciton intensity exceeds the total biexciton
intensity by a about factor of $r \approx 3$ for all considered
excitation powers (see the inset in Fig.~\ref{BiExDec}, solid
squares and circles, respectively). This fact indicates that
excitons can be formed not only as a result of biexciton radiative
decay, but also as a result of biexciton dissociation. Furthermore,
the excitation pulse can create some initial exciton population
which can be seen in Fig.~\ref{BiExDec} as a nonzero value of the
exciton intensity corresponding to the maximal biexciton intensity.
The inset in Fig.~\ref{BiExDec} demonstrates that both exciton and
biexciton time-integrated intensities show almost identical power
dependences, which at low $P$ can be approximated by $\propto
P^{1.19}$ and $\propto P^{1.12}$, respectively, with the exponent
about twice as large as that for the excitons upon exciton-resonant
excitation ($\propto P^{0.54}$, open squares in the inset, the
sublinear dependence might be related to the fact that we do not register the emission in the direction of the direct reflection of the resonant beam
which shows a large secondary emission intensity within a few picoseconds \cite{Haacke1997}). These $P$-dependences confirm the fact that biexcitons
are created by two-photon absorption process. Furthermore, the
possible initial exciton population is also created by two-photon
rather than by the standard one-photon absorption process. However,
more likely all the excitons are created from decay or dissociation
of biexcitons only, and the initial exciton population reflects the
number of excitons created from biexcitons within our time
resolution.

The dynamics of the exciton-biexciton system can be described by the
following rate equations \cite{Kim1994}:
\begin{eqnarray}
\frac{dN_\text{x}}{dt} = -\frac{N_\text{x}}{\tau_\text{x}} +
\frac{N_\text{b}}{\tau_b}+2\frac{N_\text{b}}{\tau_\text{d}} -
2F(N_\text{x}), \label{BiExFormX}
\\
\frac{dN_\text{b}}{dt} =
-\frac{N_\text{b}}{\tau_\text{b}}-\frac{N_\text{b}}{\tau_\text{d}} +
F(N_\text{x}), \label{BiExFormB}
\end{eqnarray}
where $N_\text{x}$ and $N_\text{b}$ are the exciton and biexciton
concentrations, respectively; $\tau_\text{x}$, $\tau_\text{b}$, and
$\tau_\text{d}$ are the times of the whole exciton population
radiative decay, biexciton recombination, and biexciton
dissociation, respectively; the term $F(N_\text{x})$ describes
biexciton formation from two excitons. In the description of the
fast stage we disregard the formation of secondary biexcitons
described by the last term in both equations. This process
determines the biexciton intensity when quasiequilibrium between
exciton and biexciton populations is established. This intensity is
much smaller than the biexciton intensity at the considered fast
stage for the same level of the exciton intensity (compare
Figs.~\ref{Spectra}(b) and \ref{BiExDec}).
Equations~(\ref{BiExFormX}) and (\ref{BiExFormB}) without the last term
have the following solution:
\begin{eqnarray}
N_\text{x} &=& N_\text{b}^0
\frac{2/\tau_1-1/\tau_\text{b}}{1/\tau_\text{1}-1/\tau_\text{x}}
\bigl[(1+\alpha)\text{e}^{-t/\tau_\text{x}}-\text{e}^{-t/\tau_1}\bigr],
\label{SolX}
\\
N_\text{b} &=& N_\text{b}^0
\text{e}^{-t/\tau_1},
\label{SolB}
\end{eqnarray}
where $1/\tau_1=1/\tau_\text{b}+1/\tau_\text{d}$ is the biexciton
intensity decay time, $N_\text{b}^0$ is the initial biexciton
concentration, parameter $\alpha \geq 0$ determines the initial
exciton concentration: $N_\text{x}^0=\alpha
N_\text{b}^0(2/\tau_1-1/\tau_\text{b})/(1/\tau_\text{1}-1/\tau_\text{x})$.
It is easy to show that the ratio of the total time-integrated
exciton intensity to the total biexciton intensity is
\begin{equation}
r=(2\frac{\tau_\text{b}}{\tau_1}-1)(\alpha\frac{\tau_\text{x}}{\tau_\text{x}-\tau_1}+1),
\label{Eqr}
\end{equation}
and, correspondingly, the expression for biexciton recombination
time is
\begin{equation}
\tau_\text{b}=\frac{\tau_1}{2}\bigl[1 + r
(\alpha\frac{\tau_\text{x}}{\tau_\text{x}-\tau_1}+1)^{-1}\bigr].
\label{Eqr}
\end{equation}
Thus, $\tau_\text{b}\leq \tau_1 (1+r)/2$, and since
$1/\tau_1=1/\tau_\text{b}+1/\tau_\text{d}$, $\tau_\text{b}\geq
\tau_1$. For experimental values $\tau_1=35$~ps and $r=2.9$ we have
$35 \leq \tau_\text{b}\leq 68$~ps. To get the more precise value of
$\tau_\text{b}$, we extract $\alpha=0.17$ and $\tau_\text{x}=69$~ps
(such a short $\tau_\text{x}$ is related to the low exciton
temperature in the first $100$~ps) from the fit of the exciton
intensity kinetics with Eq.~\eqref{SolX}. Finally we obtain the
biexciton recombination time $\tau_\text{b}=55$~ps and dissociation
time $\tau_\text{d}=94$~ps.

Interestingly, the biexciton recombination time
$\tau_\text{b}=55$~ps is larger than the exciton recombination time
$\tau_\text{0}=16$~ps. However, $\tau_\text{0}$ and $\tau_\text{b}$
have different physical meanings. The exciton recombination time
determines the rate of radiative decay of only low-$k$ radiative
excitons, while the total exciton population, mainly consisting of
nonradiative reservoir, decays at longer times $\tau_\text{x}\sim
400$~ps, as it follows from the experiment (Fig.~\ref{Spectra}(b))
at $t\gtrsim 200$~ps. On the other hand, biexciton radiative decay
is allowed for any biexciton wave vector since it can be transmitted
to the residual exciton. Thus, $\tau_\text{b}$ determines the
radiative decay of the whole biexciton population. The value
$\tau_\text{b}=55$~ps $> \tau_\text{0}$ is compatible with
theoretical results of Ref.~\cite{Citrin1994}, where $\tau_\text{b}$
was calculated for all
biexciton momenta. 
On the other hand, calculations performed for biexcitons with almost
zero momentum within the giant oscillator strength model
\cite{Ivanov1997} as well as four-wave mixing experiments, which
also probe almost zero-momentum biexcitons, \cite{Langbein2000} give
somewhat smaller biexciton recombination times: 20 and 11~ps,
respectively. The obtained value of $\tau_\text{b}=55$~ps is much
smaller than the biexciton recombination time evaluated in
Ref.~\cite{Kim1998} (330~ps) as the time of the biexciton intensity
decay under intense nonresonant excitation in the assumption that
the biexciton is the dominant species. As we will show later this
assumption is not valid.

\subsection{Exciton-biexciton thermodynamics}
Now, let us consider the behavior on the time scale
$t>\tau_{0},\tau_\text{b}$. One can easily estimate that, contrary to the
common belief \cite{Kim1998}, the biexciton population is much
smaller than the exciton one even for comparable exciton and
biexciton intensities $I_\text{x}\sim I_\text{b}$ :
$N_\text{b}/N_\text{x} \approx I_\text{b}\tau_\text{b}/I_\text{x}
\tau_\text{x} \sim \tau_\text{b}/\tau_\text{x}\sim 55/400 \sim 0.1$.
Furthermore, taking into account the relatively slow change of the
biexciton concentration $dN_\text{b}/dt\approx 2
N_\text{b}/t_\text{x}$ at the considered time scale
(Fig.~\ref{Spectra}), it follows from
Eqs.~\eqref{BiExFormX} and \eqref{BiExFormB} that $N_\text{b} \approx
N_\text{b}^\text{eq}/(1+\tau_\text{d}/\tau_\text{b}) \approx 0.4
N_\text{b}^\text{eq}$, where $N_\text{b}^\text{eq}=\tau_\text{d}F(N_\text{x})$ is the equilibrium biexciton concentration. Thus, the
biexciton concentration is far from the equilibrium one due to the
short radiative decay time of biexcitons compared to the biexciton
formation/dissociation time. The biexciton population is driven by
the excitons just like the nonequilibrium polariton population in a
microcavity is driven by the exciton reservoir \cite{Belykh2014}.

It is instructive to estimate the absolute value of the exciton density $N_\text{x}$. Poor knowledge of the QW absorption coefficient makes it difficult to determine $N_\text{x}$ directly from the excitation density. For our estimates, we used the exciton line blue-shift which is, according to Ref.~\cite{Schmitt-Rink1985}, $\delta E \approx 3 E_\text{xb} \lambda_\text{2D}^2 N_\text{x}$ (the detailed calculations of exciton-exciton interaction are given in Refs.~\cite{Ciuti1998,De-Leon2001}), where $E_\text{xb}$ is the exciton binding energy and $\lambda_\text{2D}$ is the two-dimensional exciton Bohr radius defined as in Ref.~\cite{Ciuti1998}. For excitation power $P=0.25$~mW (Fig.~\ref{Spectra}) at the beginning of the kinetics $\delta E \approx 0.2$~meV; taking $E_\text{xb}=7$~meV, $\lambda_\text{2D}=10$~nm, we obtain $N_\text{x}\approx 10^{10}$~cm$^{-2}$. The biexciton density can be determined from the exciton density in two ways: (i) From the ratio of the biexciton and exciton intensities as it was described above: $N_\text{b} \approx N_\text{x} I_\text{b}\tau_\text{b}/I_\text{x}
\tau_\text{x}$. For $N_\text{x} = 10^{10}$~cm$^{-2}$ and $I_\text{b}/I_\text{x}=0.4$ at the beginning of the kinetics for $P=0.25$~mW (Fig.~\ref{Spectra}) we obtain $N_\text{b}\approx 0.6 \times 10^{9}$~cm$^{-2}$. (ii) By calculating the equilibrium biexciton density $N_\text{b}^\text{eq}$ corresponding to a given exciton density from the law of mass action \cite{Kim1998}:
\begin{equation}
\frac{N_\text{x}^2}{N_\text{b}^\text{eq}}=\frac{g_\text{x}^2 m_\text{x}T}{4\pi\hbar^2}\exp(-E_\text{b}/T),
\label{EqSaha}
\end{equation}
where $m_\text{x}\approx0.3 \times 10^{-27}$~g is the two dimensional exciton effective mass \cite{Hillmer1989}, $g_\text{x}=4$ is the exciton spin degeneracy, and we put $k_\text{B}=1$. For $N_\text{x} = 10^{10}$~cm$^{-2}$, $T=10$~K we obtain $N_\text{b}^\text{eq} \approx 1.3 \times 10^9$~cm$^{-2}$. As mentioned above, due to the short biexciton lifetime, actual biexciton density $N_\text{b} \approx 0.4
N_\text{b}^\text{eq} \approx 0.5 \times 10^9$~cm$^{-2}$.
Good agreement between biexciton densities calculated in two different ways justifies our estimate for the exciton density.

Although the ratio of the total biexciton and exciton concentrations
is far from equilibrium, both exciton and biexciton momentum
distributions should be close to the thermal distribution
characterized by the internal temperature $T$ in general not equal
to the lattice temperature $T_\text{latt}$. This is true since the
thermalization time determined by interparticle scattering at
considered excitation densities is shorter than $\tau_\text{0}$ and
$\tau_\text{b}$ \cite{Knox1986, Knox1988} (interparticle scattering time can be estimated from the homogeneous density-dependent broadening which is up to 1~meV in our experiments). Short thermalization time also excludes
the effects related to the laser-mediated coherence which may be important only at low densities \cite{Haacke1997, Garro1999, Langbein2000a, Hayes2000, Savona2000, Kocherscheidt2003, Savona2002}. Furthermore, the estimated exciton density indicates that
the exciton ensemble is nondegenerate at the considered temperatures and can be described by the classical Boltzmann distribution.

\subsection{Internal temperature dynamics}
Now, we will describe how the internal temperature $T$ of the
exciton-biexciton system can be determined from the decay time of
the exciton population $\tau_\text{x}$, and discuss the dynamics of
$T$.

The radiative decay rate of the whole exciton population is given by
the following relation:
$N_\text{x}/\tau_\text{x}=N_\text{rad}/\tau_\text{0}$, where
$N_\text{rad}$ is the concentration of excitons in the radiative
zone with wave vectors within the light cone $k < \omega n/c$ and spins
$\pm 1$. This concentration can be evaluated as $N_\text{rad}\approx
\delta E_\text{r}(1/2)f(0)D=\delta E_\text{r}N_\text{x}/2T$, where
we put the energy of the bottom of the exciton dispersion curve to
zero, $\delta E_\text{r} = (\hbar\omega
n/c)^2/2m_\text{x} \approx 80$~$\mu$eV is the maximal energy of the
radiative excitons, $f(E)=(\pi\hbar^2
N_\text{x}/2m_\text{x}T)\exp(-E/T)$ is the occupancy of exciton
states given by the Boltzmann distribution, $D=2
m_\text{x}/\pi\hbar^2$ is the exciton density of states per unit
area, $n\approx 3.3$ is the refractive index of the GaAs, and we
used $T\gg \delta E_\text{r}$. Finally, for the radiative decay time
of the whole exciton population we have
\begin{equation}
\tau_\text{x}=2\tau_\text{0}T/\delta E_\text{r}=\frac{T}{A},
\label{tauT}
\end{equation}
and we introduced phenomenological constant $A=\delta
E_\text{r}/2\tau_\text{0}$ which will be determined in the
experiment and related to the measured $\tau_\text{0}$.

\begin{figure}
\begin{center}
\includegraphics[width=\columnwidth]{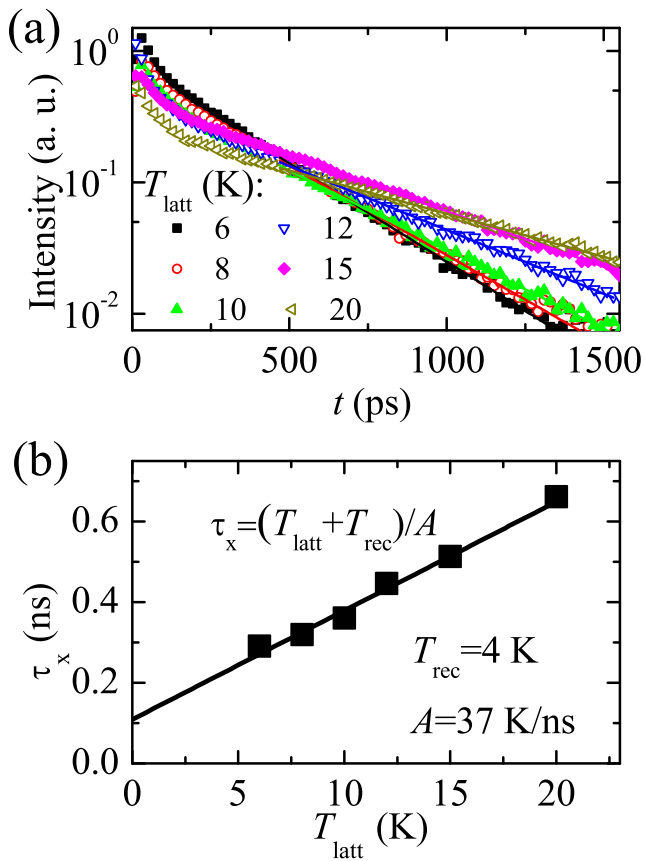}
\caption{(a) Dynamics of the exciton PL intensities (symbols) at different
lattice temperatures for exciton-resonant excitation with $P =
0.13$~mW. The lines show the exponential fits. (b) Lattice temperature dependence of the exciton PL decay
time (symbols). The line shows the linear fit.} \label{FigTDep}
\end{center}
\end{figure}

Figure \ref{FigTDep}(a) shows the exciton PL kinetics at different
lattice temperatures. At long times, the PL decay can be described
by a monoexponential function and the decay time is equal
$\tau_\text{x}$ if we neglect
biexciton formation. 
The dependence of $\tau_\text{x}$ on $T_\text{latt}$ is shown in
Fig.~\ref{FigTDep}(b). As expected, this dependence is linear
$\tau_\text{x}=(T_\text{latt}+T_\text{rec})/A$, however, there is
nonzero offset suggesting that the internal exciton temperature is
higher than lattice temperature by a constant value
$T_\text{rec}\approx 4$~K. The origin of $T_\text{rec}$ will be
discussed in the following. The experimental value $A=37$~K/ns corresponds to
the exciton recombination time $\tau_\text{0}\approx 13$~ps in  good
agreement with measured value of $16$~ps. The linear dependence of
the exciton population decay time on temperature was also observed
in previous experiments [Ref.~\cite{Feldmann1987} ($A\approx
20$~K/ns for a 4-nm QW) and Ref.~\cite{Martinez-Pastor1993}
($A\approx 45$~K/ns for a 4-nm QW)] and calculated theoretically
[Ref.~\cite{Andreani1991} ($A\approx 29$~K/ns for a 10-nm QW)].

We have shown that the exciton internal temperature is proportional
to the decay time of the exciton intensity (Eq.~\eqref{tauT}).
However, in the experiment the decay time can be defined
unambiguously only at large times, when the intensity decay is
exponential and the internal temperature almost reached its
steady-state value. Now, we extract the internal temperature at any
time $T(t)$ from the dynamics of the exciton and biexciton
intensities. The exciton intensity (defined as the number of particles emitted in the unit time)
$I_\text{x}(t)=N_\text{x}(t)/\tau_\text{x}(t)=A N_\text{x}(t) /
T(t)$. On the other hand, $N_\text{x}(t) = \int_{t}^{\infty}
(I_\text{x}(t') + I_\text{b}(t')) dt'$, where we use $N_\text{b} \ll
N_\text{x}$. Thus,
\begin{equation}
T(t) = A \frac{\int_{t}^{\infty} (I_\text{x}(t') + I_\text{b}(t'))
dt'}{I_\text{x}(t)}. \label{EqTI}
\end{equation}
This expression offers a method to determine the exciton
temperature from the PL dynamics. In the case of monoexponential
decay and when $I_\text{b}(t) \ll I_\text{x}(t)$, Eq.~\eqref{EqTI} is
reduced to Eq.~\eqref{tauT}.

\begin{figure}
\begin{center}
\includegraphics[width=1\columnwidth]{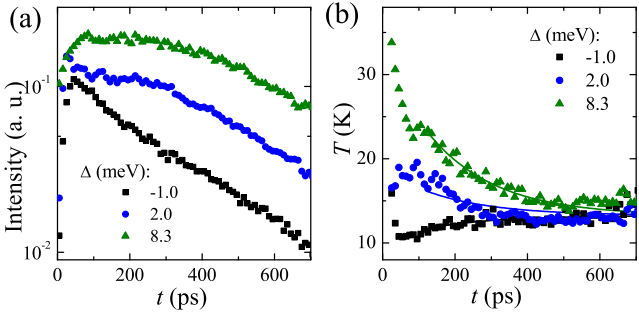}
\caption{Dynamics of the exciton intensity (a) and internal
temperature (b) for excitations with different excess energies
$\Delta$ with respect to the exciton resonance and $P = 0.13$~mW.
Lines in Fig.~\ref{FigTDyn}(b) show exponential fits.
$T_\text{latt}=10$~K.} \label{FigTDyn}
\end{center}
\end{figure}

To test the validity of this method, we perform experiments using
excitations with different excess energies
$\Delta=E_\text{laser}-E_\text{x}$ with respect to the exciton
resonance, which defines the initial exciton temperature. The
excitation density is relatively low, so the biexciton recombination
heating (see following) is small (for $\Delta < 0$ it is small at $t \gtrsim
\tau_\text{b}$). The exciton PL dynamics under excitation with
different excess energies is shown in Fig.~\ref{FigTDyn}(a).
Figure~\ref{FigTDyn}(b) shows the exciton temperature dynamics
determined according to Eq.~\eqref{EqTI} from the measured exciton
and biexciton intensity dynamics. As expected, the temperature
relaxes from the value determined by the excess energy and reaches
the steady-state value $T_\text{st} = T_\text{latt}+T_\text{rec}$ at
long times. The following qualitative criterion applies: the faster
the intensity decrease, the lower the temperature and vice
versa [compare curves in Figs.~\ref{FigTDyn}(a) and
\ref{FigTDyn}(b)]. Our method can be applied even for low exciton
temperatures and concentrations, where the temperature can not be
determined from the Boltzmann tail in the PL spectra, which
originates either from the e-h plasma recombination
\cite{Szczytko2004} or from the longitudinal optical (LO) phonon replica of exciton
recombination \cite{Bieker2014}.

Now, we show that increase of the internal exciton temperature $T$ by
$T_\text{rec}$ [offset of the dependence in Fig.~\ref{FigTDep}(b)]
is related to recombination heating \cite{Bimberg1985} (evaporative
optical heating \cite{Ivanov2004}), a process similar to evaporative
cooling in atomic systems \cite{Cornell2002, Davis1995}. The average
exciton energy is $T$, but only excitons with low energies $< \delta
E_\text{r} \ll T$ recombine; as a result, the average exciton energy
is increased. Here, we disregard biexciton formation, which is
insignificant at long times. To describe exciton recombination
heating quantitatively, let us consider the evolution of the exciton
system total energy per unit area $N_\text{x}T$:
\begin{equation}
\frac{d}{dt}(N_\text{x}T)=-\frac{N_\text{x}}{\tau_\text{x}}\frac{\delta
E_\text{r}}{2}-\kappa (T-T_\text{latt})N_\text{x}. \label{EqEDyn}
\end{equation}
The total energy changes due to the radiative decay (first term, where $E_\text{r}/2$ is the average energy
of recombining excitons) and
due to the emission/absorption of phonons (second term). The latter
process tends to equalize the temperatures of the exciton system and
the lattice, so in the first-order approximation, its rate is
proportional to $T-T_\text{latt}$. The proportionality coefficient
$\kappa$ characterizes the strength of the exciton-phonon
interaction. Taking into account that
$dN_\text{x}/dt=-N_\text{x}/\tau_\text{x}=-A N_\text{x}/T$ and
$\delta E_\text{r} \ll T$, we obtain the solution of
Eq.~\eqref{EqEDyn}:
\begin{equation}
T=T_0 e^{-\kappa t} + (T_\text{latt}+\frac{A}{\kappa})(1-e^{-\kappa
t}), \label{EqTDyn}
\end{equation}
where $T_0$ is the initial temperature. Thus, the steady-state value
of the exciton temperature achieved at $t\gg 1/\kappa$
\begin{equation}
T_\text{st}=T_\text{latt}+\frac{A}{\kappa}
\end{equation}
is higher than $T_\text{latt}$ by a constant
$T_\text{rec}=A/\kappa$, in agreement with the results in
Figs.~\ref{FigTDep}(b) and \ref{FigTDyn}(b). The recombination
heating temperature
\begin{equation}
T_\text{rec}=\frac{A}{\kappa}=\frac{\delta E_\text{r}}{2 \tau_0
\kappa} \label{EqTRad}
\end{equation}
is determined by the exciton recombination rate and exciton-phonon
interaction constant. Interestingly, only interaction with the
lattice prevents the temperature from rising infinitely. By fitting
the temperature dynamics in Fig.~\ref{FigTDyn}(b) with
Eq.~\eqref{EqTDyn}, we determine $\kappa\approx 5.4$~ns$^{-1}$
(exciton-phonon scattering time $1/\kappa\approx 180$~ps), and we
can calculate the recombination heating temperature
$T_\text{rec}=A/\kappa \approx 37 / 5.4 \approx 6.9$~K which is in a
reasonable agreement with the value $T_\text{rec}=4$~K determined
from the data in Fig.~\ref{FigTDep}(b).

Another process that, in addition to exciton recombination heating,
contributes to the increase in the exciton temperature is the
formation of biexcitons with their \emph{subsequent recombination}.
Indeed, each biexciton formation event with subsequent recombination
releases energy $E_\text{b}$ for the exciton system, while the
emitted photon has energy $-E_\text{b}$. Thus, taking into account
that $N_\text{b} \ll N_\text{x}$, while intensities $I_\text{b} =
N_\text{b} / \tau_\text{b}$ and $I_\text{x} = N_\text{x} /
\tau_\text{x}$ might be comparable, one has to add a term
$E_\text{b} N_\text{b} / \tau_\text{b}$ to the right-hand side of
Eq.~\eqref{EqEDyn}. This term describes the energy increase due to
biexciton recombination. As a result, the temperature dynamics is
described by the equation
\begin{equation}
\frac{dT}{dt}=A-\kappa (T-T_\text{latt}) +
\frac{E_\text{b}}{\tau_\text{b}}\frac{N_\text{b}}{N_\text{x}},
\label{EqTDynBiEx}
\end{equation}
leading to
\begin{multline}
T=T_0 e^{-\kappa t} + (T_\text{latt}+\frac{A}{\kappa})(1 -
e^{-\kappa t})
\\
+\frac{E_\text{b}}{\tau_\text{b}}\int_{0}^{t}\frac{N_\text{b}(t')}{N_\text{x}(t')}e^{-\kappa(t-t')}dt'
\label{EqTDynBiEx}
\end{multline}
The last term determines the temperature increase due to biexciton
recombination. To calculate this term, one should obtain the time
dependence of $N_\text{x}$ and $N_\text{b}$ from
Eqs.~(\ref{BiExFormX}) and (\ref{BiExFormB}). However, the constants
$\tau_\text{x}$, $\tau_\text{d}$ and formation rate $F(N_\text{x})$
in these equations in turn depend on $T$. To get an approximate
expression for the last term in Eq.~\eqref{EqTDynBiEx}, we put
$N_\text{x}\propto \exp(-t/\tau_\text{x})$ and $N_\text{b}\propto
\exp(-2t/\tau_\text{x})$ and disregard the dependence of
$\tau_\text{x}$ on $T$. Finally,
\begin{multline}
T=T_0 e^{-\kappa t} + (T_\text{latt}+\frac{A}{\kappa})(1 -
e^{-\kappa t})
\\
+\frac{E_\text{b}}{\tau_\text{b}\kappa -
\tau_\text{b}/\tau_\text{x}}\frac{I_\text{b}(0)}{I_\text{x}(0)}(e^{-t/\tau_\text{x}}
- e^{-\kappa t}). \label{EqTDynBiEx1}
\end{multline}
Thus, biexciton recombination can increase the internal temperature
by as much as $\sim E_\text{b}$.

\begin{figure}
\begin{center}
\includegraphics[width=\columnwidth]{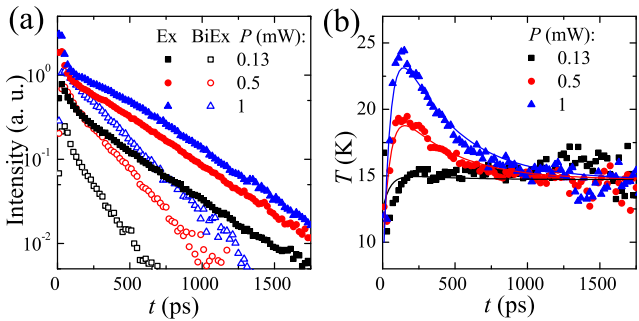}
\caption{Dynamics of the exciton (solid symbols) and biexciton (open
symbols) intensities (a) and the internal temperature (b) for
exciton-resonant excitation ($\Delta = 0$) with different powers.
Lines in Fig.~\ref{FigTDynBiEx}(b) show biexponential fits.
$T_\text{latt}=10$ K.} \label{FigTDynBiEx}
\end{center}
\end{figure}

Figure~\ref{FigTDynBiEx}(b) shows the exciton temperature dynamics
for different powers of exciton-resonant excitation. The
corresponding exciton and biexciton intensities are shown in
Fig.~\ref{FigTDynBiEx}(a). For the lowest excitation power, the
temperature monotonously increases to the steady-state value
$T_\text{st}$. On the other hand, for higher powers, when the
biexciton intensity becomes comparable to the exciton intensity
(Fig.~\ref{FigTDynBiEx}(a)), the temperature first increases above
$T_\text{st}$ and then decreases to $T_\text{st}$. This initial
temperature rise results from biexciton recombination heating and is
proportional to $I_\text{b}/I_\text{x}$. As the biexciton intensity
becomes much lower than the exciton intensity, the temperature
relaxes to $T_\text{st}$. The dependences in
Fig.~\ref{FigTDynBiEx}(b) are fitted with double-exponential
functions in accordance with Eq.~\eqref{EqTDynBiEx1} with the
temperature decay time set to $\tau_\text{x}= 0.38$~ns.

The described heating mechanisms cancel out the possibility of
exciton BEC in the considered system even for resonant excitation.
Indeed, the system temperature can not be lower than $\sim 4$~K for a
time longer than few tens of picoseconds even for $T_\text{latt} =
0$. This already gives a BEC threshold density of at least
$N_\text{thr} \sim 2 m_\text{x}T_\text{rec}/\pi \approx 1.6\times
10^{12}$~cm$^{-2}$ which is of the same level as the exciton
saturation density $N_\text{sat}\sim 1/a_\text{B}^2 \sim
10^{12}$~cm$^{-2}$. Furthermore, an increase in the exciton density
leads to a proportional increase in the temperature due to biexciton
recombination heating $T_\text{b} \propto N_\text{b}/N_\text{x}
\propto N_\text{x}$. Experiment shows that even a moderate increase
in $N_\text{x}$, to the level when
$I_\text{b}$ becomes comparable to $I_\text{x}$, 
leads to an increase in $T$ up to $\sim 10$~K. This effect forbids
exciton BEC even if one forgets about exciton saturation.

\section{Conclusion}
A consistent picture of exciton and biexciton recombination in
GaAs/AlGaAs quantum wells has been developed. The exciton
recombination time has been determined to be $\tau_\text{0}=16$~ps,
while the radiative decay time of the whole exciton population
including large-momentum nonradiative part is $\tau_\text{x} = T/(37
$ K/ns$) \gg \tau_\text{0}$. The biexciton recombination time, which
also characterizes the radiative decay of the whole biexciton
population, has been determined to be $\tau_\text{b} =55$~ps. As a
result of $\tau_\text{b} \ll \tau_\text{x}$, the biexciton
concentration is significantly lower than one would expect in
thermal equilibrium. It is also much lower than the exciton
concentration, while the exciton and biexciton emission intensities
can be comparable. Thus, the main role of the biexciton states in
the thermodynamics of the whole system is that they provide an
additional radiative channel for excitons.

A method of determining the temperature dynamics of the exciton
system has been developed. It is based on the proportionality of the
radiative decay time of the whole exciton population and temperature
of the exciton system and also takes into account biexciton
recombination channel. The method has been tested in experiments
with different excess energies of the excitation photons which
define the initial temperatures of the system.

The heating of the exciton system by exciton recombination has been
analyzed theoretically and revealed in the experiment. This effect
is analogous to evaporative cooling in atomic systems. An increase
in the temperature due to recombination heating ($\sim 4$~K in the
QWs under study) is proportional to the exciton recombination rate
and inversely proportional to the rate of cooling by acoustic
phonons.

If has been shown that biexciton formation and subsequent
recombination also lead to an increase in the exciton temperature by
an amount proportional to the ratio of the biexciton and exciton
emission intensities. For comparable biexciton and exciton
intensities, this increase can be as large as tens of Kelvins.

The revealed recombination heating effects imply new restrictions on
the Bose-Einstein condensation of excitons. These restrictions
cancel out BEC in many exciton systems even where the thermalization
rate exceeds the radiative decay rate (like direct excitons in
GaAs/AlGaAs QWs) and even for resonantly created cold exciton gas.

The main quantitative conclusions of the paper are summarized in
Table~\ref{Tab}.

\begin{table*}
\begin{center}
\begin{tabular}{ l  l }
    \hline \hline
    Exciton recombination time & $\tau_\text{0} = 16$~ps \\
    Biexciton recombination time & $\tau_\text{b} = 55$~ps \\
    Radiative decay time
    of the whole exciton population  & $\tau_\text{x} = T/A$, $A=37$~K/ns \\
    Exciton temperature dynamics & $T(t) = A [\int_{t}^{\infty} (I_\text{x}(t') + I_\text{b}(t'))
dt']/I_\text{x}(t)$ \\
    Exciton recombination heating temperature & $T_\text{rec}=A/\kappa \approx
    4$~K \\
    Biexciton recombination heating temperature &
    $T_\text{b}=(E_\text{b}/\tau_\text{b})\int_{0}^{t}[N_\text{b}(t')/N_\text{x}(t')]\exp[-\kappa(t-t')]dt'$
    \\
    \hline \hline
    \end{tabular}

    \caption{Summary of the main quantitative results.}
\label{Tab}
\end{center}
\end{table*}

\begin{acknowledgements}
We are grateful to M.~L.~Skorikov for careful reading of the
manuscript, valuable remarks and advices, and to N.~N.~Sibeldin and
V.~A.~Tsvetkov for useful discussions. The work is supported by the
Russian Science Foundation (Project No.~14-12-01425).
\end{acknowledgements}

\newpage

\end{document}